\documentclass[aps, prb, reprint, showpacs, showkeys, onecolumn, 11pt, superscriptaddress, notitlepage]{revtex4-1}
\usepackage{graphicx} 

\begin{document}

\bibliographystyle{unsrt}

\title{Charge noise and spin noise in a semiconductor quantum device}

\author{Andreas V. Kuhlmann}
\affiliation{Department of Physics, University of Basel, Klingelbergstrasse 82, CH-4056 Basel, Switzerland}

\author{Julien Houel}
\affiliation{Department of Physics, University of Basel, Klingelbergstrasse 82, CH-4056 Basel, Switzerland}

\author{Arne Ludwig}
\affiliation{Department of Physics, University of Basel, Klingelbergstrasse 82, CH-4056 Basel, Switzerland}
\affiliation{Lehrstuhl f\"{u}r Angewandte Festk\"{o}rperphysik, Ruhr-Universit\"{a}t Bochum, D-44780 Bochum, Germany}

\author{Lukas Greuter}
\affiliation{Department of Physics, University of Basel, Klingelbergstrasse 82, CH-4056 Basel, Switzerland}

\author{Dirk Reuter}
\affiliation{Lehrstuhl f\"{u}r Angewandte Festk\"{o}rperphysik, Ruhr-Universit\"{a}t Bochum, D-44780 Bochum, Germany}
\affiliation{Department Physik, Universit\"{a}t Paderborn, Warburger Strasse 100, D-33098 Paderborn, Germany}

\author{Andreas D. Wieck}
\affiliation{Lehrstuhl f\"{u}r Angewandte Festk\"{o}rperphysik, Ruhr-Universit\"{a}t Bochum, D-44780 Bochum, Germany}

\author{Martino Poggio}
\affiliation{Department of Physics, University of Basel, Klingelbergstrasse 82, CH-4056 Basel, Switzerland}

\author{Richard J. Warburton}
\affiliation{Department of Physics, University of Basel, Klingelbergstrasse 82, CH-4056 Basel, Switzerland}

\date{\today}

\begin{abstract}
Solid-state systems which mimic two-level atoms are being actively developed. Improving the quantum coherence of these systems, for instance spin qubits or single photon emitters using semiconductor quantum dots, involves dealing with noise. The sources of noise are inherent to the semiconductor and are complex. Charge noise results in a fluctuating electric field, spin noise in a fluctuating magnetic field at the location of the qubit, and both can lead to dephasing and decoherence of optical and spin states. We investigate noise in an ultra-pure semiconductor using a minimally-invasive, ultra-sensitive, local probe: resonance fluorescence from a single quantum dot. We distinguish between charge noise and spin noise via a crucial difference in their optical signatures. Noise spectra for both electric and magnetic fields are derived. The noise spectrum of the charge noise can be fully described by the fluctuations in an ensemble of localized charge defects in the semiconductor. We  demonstrate the ``semiconductor vacuum" for the optical transition at frequencies above 50 kHz: by operating the device at high enough frequencies, we demonstrate transform-limited quantum dot optical linewidths.
\end{abstract}

\maketitle

Semiconductor quantum dots are hosts for spin qubits\cite{Loss1998,Petta2005}. Optically-active quantum dots, for instance self-assembled quantum dots, are in addition to spin qubits potentially excellent single photon sources\cite{Shields2007}. Optimizing performance demands an understanding of noise and a strategy to circumvent its deleterious effects\cite{Fischer2009b}. There are two main sources of noise in a semiconductor. Charge noise arises from occupation fluctuations of the available states and leads to fluctuations in the local electric field. This results in shifts in the optical transition energy of a quantum dot via the dc Stark effect and is one mechanism by which the optical linewidth of a self-assembled quantum dot can be significantly increased above the transform limit \cite{Hogele2004,Atature2006,Houel2012}. Charge noise can also result in spin dephasing via the spin-orbit interaction, and, in particular for hole spins, via the electric field dependence of the g-factor \cite{Klotz2010,Pingenot2011}. The second source of noise, spin noise, arises typically from fluctuations in the nuclear spins of the host material and, on account of the hyperfine interaction, results in a fluctuating magnetic field (the Overhauser field) experienced by an electron spin\cite{Merkulov2002,Khaetskii2002}. Spin noise from noisy nuclei results in rapid spin dephasing in a GaAs quantum dot\cite{Greilich2006a,Xu2008,Press2010}. 

Strategies for reducing noise involve working with ultra-clean materials to minimize charge noise, and possibly nuclear spin-free materials to eliminate spin noise. Two spectacular examples are isotopically-pure diamond\cite{Balasubramanian2009} and isotopically-pure silicon\cite{Steger2012} for both of which spin coherence of localized spins is impressive. Abandoning GaAs comes however with a significant loss of flexibility for both spin qubits and quantum photonics applications. A second powerful paradigm is the use of dynamic decoupling, schemes which employ complex echo-like sequences to ``protect" the qubit from environmental fluctuations\cite{Barthel2010,deLange2010,Bluhm2011}. In this case, it is absolutely crucial for success that the noise power decreases with increasing frequency. 

For quantum dot-based single photon sources, the linewidths are in the best case (high quality material with resonant excitation) typically about a factor of two larger than the transform limit in which the linewidth is determined only by the radiative decay time\cite{Hogele2004,Atature2006,Houel2012}. This is a poor state of affairs for applications which rely on photon indistinguishability, the resource underpinning a quantum repeater for instance. It has been surmised that the increase in linewidth above the ideal limit arises from a spectral wandering \cite{Hogele2004,Houel2012} but the exact origin of the noise and its frequency dependence has not been pinned down. This ignorance makes engineering better quantum devices difficult. It is clear that untreated noisy nuclei limit the electron spin coherence of a spin qubit\cite{Merkulov2002,Khaetskii2002,Coish2004}. However, the mesoscopic nature -- a quantum dot contains $10^{5} - 10^{6}$ nuclear spins -- allows the nuclear spins to be manipulated, both quietened down and polarized\cite{Urbaszek2013}. Typically, the total nuclear spin noise is inferred indirectly from the electron spin coherence. A spin noise spectrum has been deduced at high frequencies from the time dependence of spin qubit dynamic decoupling\cite{Medford2012}, and at low frequencies from successive spin qubit readout operations\cite{Reilly2008}, leaving a gap at intermediate frequencies\cite{Fink2013}. 

We present here an investigation of noise in an ultra-clean semiconductor quantum device, using a minimally-invasive, ultra-sensitive, local probe: resonance fluorescence from a single quantum dot, Fig.\ 1(a). We present noise spectra with 4 decades of resolution in the noise amplitude (8 in the noise power) over 6 decades of frequency, from 0.1 Hz to 100 kHz, Fig.\ 2 (a),(b). Significantly, we have discovered a spectroscopic way to distinguish charge noise from spin noise, Fig.\ 3. We find that the charge noise is concentrated at low frequencies and gives a large noise amplitude but only in a small bandwidth. The spin noise lies at higher frequencies and gives much weaker noise amplitudes but over a much larger bandwidth. Remarkably, our experiment is able to reveal the full spectrum of the fluctuating nuclear spin ensemble. We translate the resonance fluorescence noise spectrum into two separate noise spectra, one for the local electric field (charge noise) and one for the local magnetic field (spin noise). Both noise sources have Lorentzian power spectra with a $1/f^{2}$-dependence at ``high" frequency. The combined noise falls rapidly with frequency becoming insignificant above 50 kHz, the ``semiconductor vacuum", for the optical transition.
\begin{figure}[t]  
\center
\includegraphics[width=\linewidth]{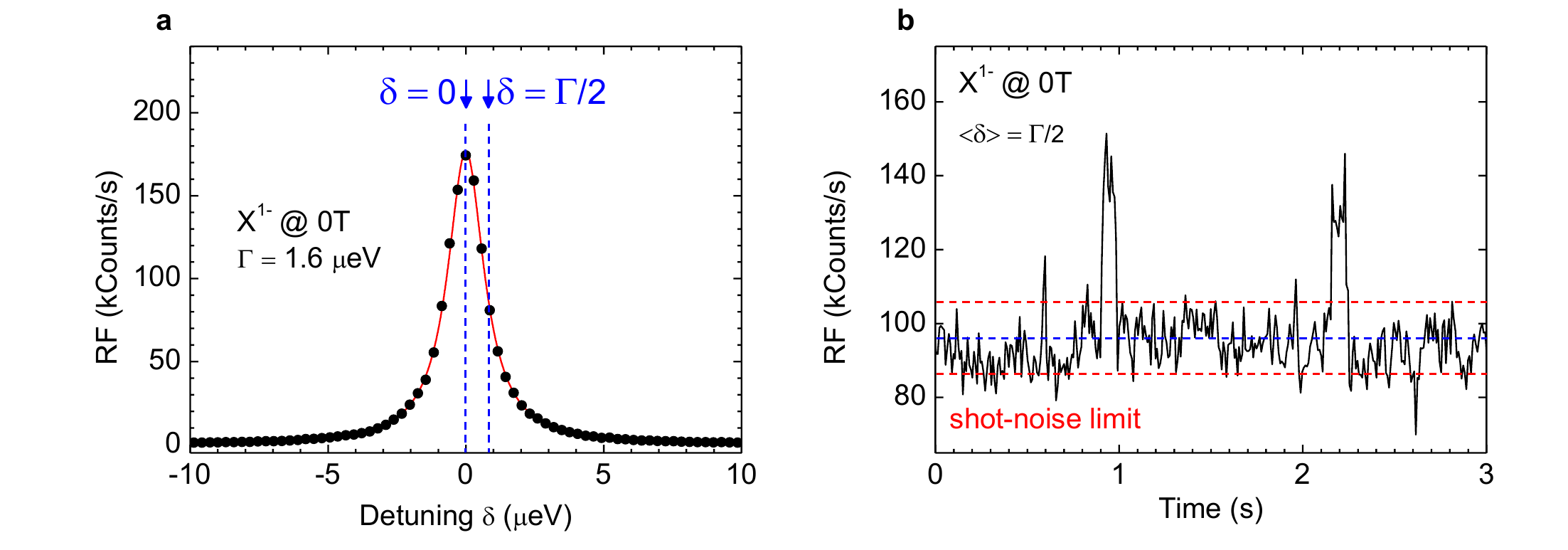}
\caption{{\bf Resonance fluorescence (RF) on a single quantum dot} (a) RF recorded on a single InGaAs quantum dot at wavelength 950.61 nm at a power corresponding to a Rabi energy of 0.55 $\mu$eV at a temperature of 4.2 K without external magnetic field. The RF was detected with a silicon avalanche photo-diode operating in single photon mode; the detuning was achieved by sweeping the gate voltage with respect to the laser using the dc Stark effect. In this case, the integration time per point was 100 ms. The solid line is a Lorentzian fit to the data with linewidth $\Gamma=1.6$ $\mu$eV (390 MHz). (b) A time-trace of the RF recorded with detuning set to half the linewidth, $\langle\delta\rangle=\Gamma/2$. The arrival time of each detected photon is stored allowing a time trace to be constructed post-experiment with an arbitrary binning time. An example is shown using a binning time of 10 ms.}
\end{figure}
The sample consists of a layer of self-assembled InGaAs quantum dots in an ultra-clean GaAs n-i-Schottky field effect device\cite{Drexler1994,Warburton2000}. A single quantum dot is driven resonantly with a coherent continuous wave laser in the linear regime. The resonance fluorescence (RF) is detected by counting individual photons. Detuning of the  quantum dot relative to the constant frequency laser is achieved by tuning the quantum dot via the dc Stark effect. Once the parameters have been set for a particular experiment, we record the arrival time of every photon for the duration of the experiment. Post-measurement, the data sets are analysed by setting a binning time and performing a fast Fourier transform to yield RF noise versus frequency $f$ spectra. 
\begin{figure}[b]  
\center
\includegraphics[width=\linewidth]{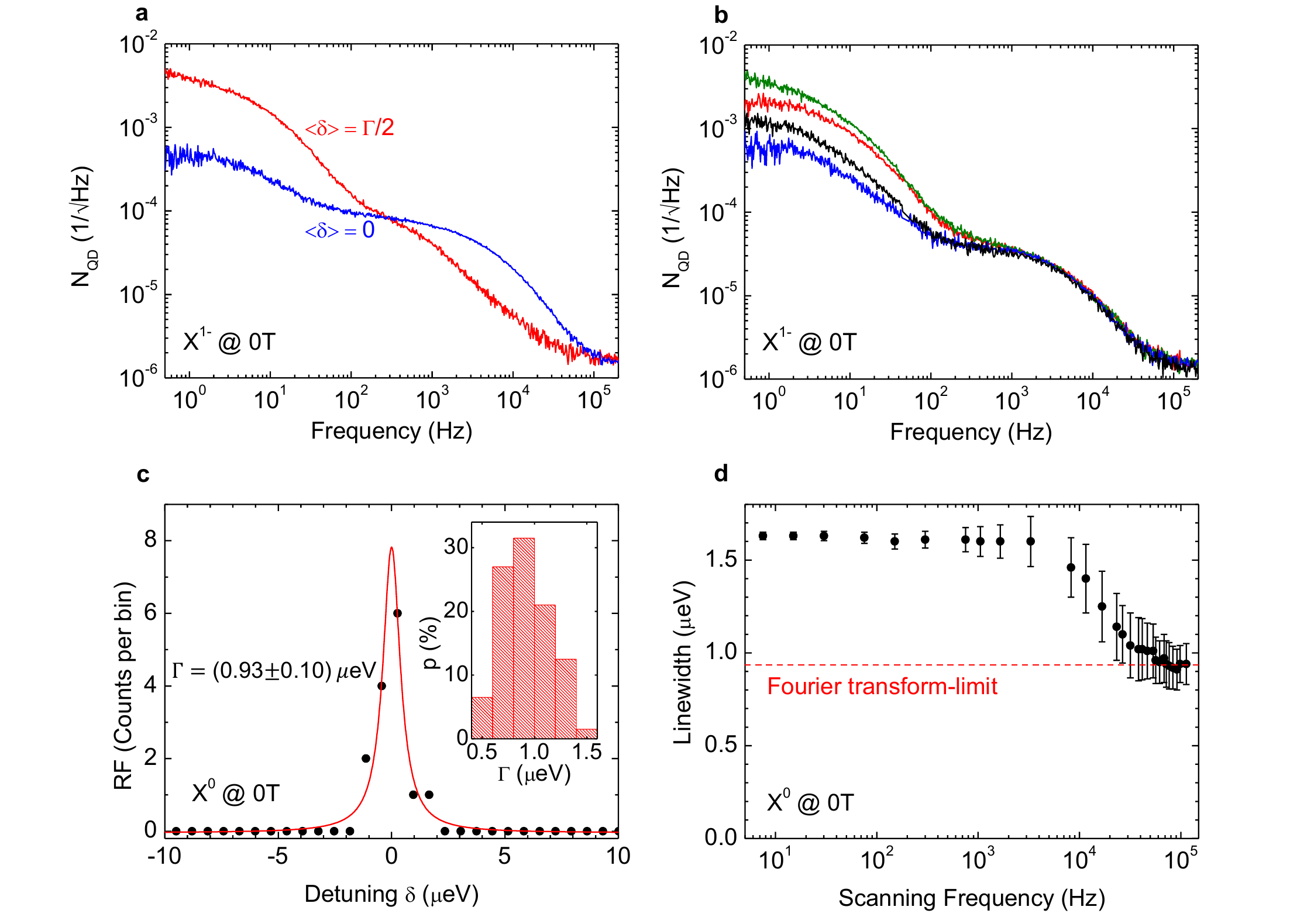}
\caption{{\bf Resonance fluorescence noise} (a) RF noise spectra recorded on a quantum dot (the one from Fig.\ 1) occupied with a single electron, the trion X$^{1-}$, for average detuning equal to zero, $\langle\delta\rangle=0$ (blue), and for $\langle\delta\rangle=\Gamma/2$ (red) at 4.2 K and $B_{\rm ext}=0$. Following the scheme in Fig.\ 3, the noise at low frequencies is shown to originate from charge noise, that at high frequencies from spin noise. Plotted is the noise spectrum of the normalized RF, $S(t)/\langle S(t)\rangle$, where $S(t)$ is the RF signal, $\langle S(t)\rangle$  the average RF signal, corrected for external sources of noise (see Methods). (b) RF noise spectra recorded on X$^{1-}$ with $\langle\delta\rangle=0$ under identical experimental conditions (4.2 K, $B_{\rm ext}=0$) in the course of the experiment. The charge noise at low frequency depends on the sample history; the spin noise at high frequency does not. (c) An example X$^{0}$ RF spectrum measured with $f_{\rm scan}=58$ kHz and t$_{\rm bin}=13$ $\mu$s. The scanning frequency is defined as $d \delta/dt/\Gamma_{0}$ where $\Gamma_{0}$ is the transform-limited linewidth. Inset: histogram of 200 linewidths recorded also with $f_{\rm scan}=58$ kHz. (d) RF linewidth against scanning frequency. The radiative lifetime is $\tau_r=(700 \pm 50)$ ps. $\Gamma$ approaches $\Gamma_{0}$ for scanning frequencies above 50 kHz.}
\end{figure}

A typical time trace of the RF is shown in Fig.\ 1(b) with binning time 10 ms. At first sight, one might think that the time trace is unlikely to be very revealing about the environmental noise at the location of the quantum dot as the experiment itself and not just the quantum dot is a source of noise, mostly shot noise. However, this experimental noise is highly reproducible. We record its spectrum carefully and, using a new protocol (see Methods) subtract it from the total noise to determine the quantum dot noise. In this way, our experiment becomes sensitive to quantum dot noise which is just a few \% of the shot noise. Our protocol (see Methods) yields a noise amplitude for the normalized RF signal $N_{\rm QD}(f)=|{\rm FT}[S_{\rm QD}(t)/\langle S(t)\rangle]|$ where ${\rm FT}[S_{\rm QD}(t)]$ is the Fourier transform of the quantum dot RF signal and $\langle S(t)\rangle$ is the average count rate over the experiment.\\

\noindent{\bf Charge noise versus spin noise} The spectrum of the noise in the RF arising from the quantum dot is shown in Fig.\ 2(a). In this case, the gate voltage $V_{g}$ is set so that the quantum dot contains a single electron and the laser drives the trion resonance, X$^{1-}$. Two features can be made out in the noise spectrum, a roll-off-like spectrum with ``high" amplitude and ``low" characteristic frequency, and a roll-off-like spectrum with ``low" amplitude and ``high" characteristic frequency. This points to the presence of two noise sources in the semiconductor. But is the noise charge noise or spin noise?
\begin{figure}[b]  
\center
\includegraphics[width=\linewidth]{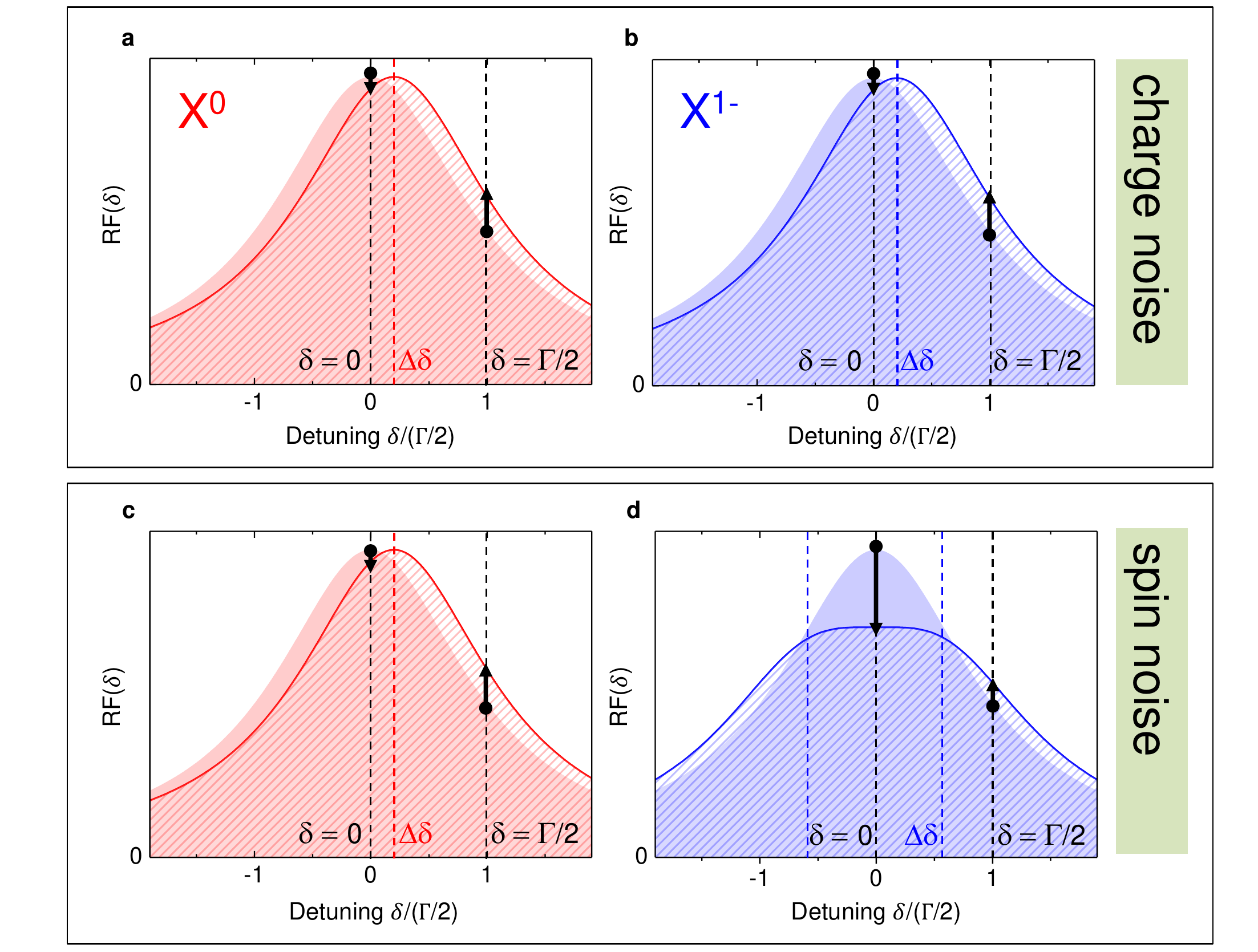}
\caption{{\bf Distinguishing between charge noise and spin noise} (a)-(d) Schematic showing the effect of charge noise and spin noise on the neutral, X$^{0}$, and charged, X$^{1-}$, excitons (applied magnetic field zero). Charge noise (noise in the local electric field) results in a ``rigid" shift of the optical resonance leading to a small change in resonance fluorescence (RF) for zero detuning $\delta=0$ and a large change in RF at $\delta=\Gamma/2$. This applies for both X$^{0}$ and X$^{1-}$, (a), (b). Without an external magnetic field, spin noise (noise in the local magnetic field experienced by a conduction electron) results in a small shift in the X$^{0}$ resonance position, qualitatively as for charge noise, (c). (d) For X$^{1-}$ however, spin noise induces a Zeeman splitting in the resonance resulting in a large change in RF at $\delta=0$ and a small change in RF at $\delta=\Gamma/2$ (zero for $\delta=\Gamma/2\sqrt{3}$), opposite to charge noise. This difference, a ``rigid" shift of the X$^{1-}$ resonance from charge noise, a ``breathing motion" in the X$^{1-}$ resonance from spin noise, allows charge noise and spin noise to be identified.}
\end{figure}
To identify the two noise sources, we present noise spectra taken with two detunings, one with average detuning zero $\langle \delta\rangle=0$ (i.e.\ detuning averaged over the experiment is zero), the other with average detuning half a linewidth, $\langle \delta\rangle=\Gamma/2$, Fig.\ 2(a). Switching from $\langle \delta\rangle=0$ to $\langle \delta\rangle=\Gamma/2$ causes the amplitude of the low frequency noise component to increase by an order of magnitude yet the amplitude of the high frequency noise component to decrease, by about a factor of five at a few kHz, Fig.\ 2(a). This crucial information allows the nature of the noise, charge or spin, to be identified.

As the local electric field $F$ fluctuates, the detuning $\delta$ of the quantum dot optical resonance with respect to the constant laser frequency fluctuates on account of the dc Stark effect. For small electric field fluctuations, the Stark shift is linear: the optical resonance shifts rigidly backwards and forwards on the detuning axis, as shown in Fig.\ 3(a),(b). The response in the RF to charge noise has a first order component in electric field for $\delta=\Gamma/2$ giving rise to large changes in the RF. Conversely, for $\delta=0$ the first order component vanishes and the change in RF is therefore smaller. Sensitivity to charge noise in the RF is therefore weak for $\langle \delta\rangle=0$ yet strong for $\langle \delta\rangle=\Gamma/2$. Spin noise results in a complementary behaviour in the absence of an external magnetic field, $B_{\rm ext}=0$. Fluctuations in the local magnetic field $B_{N}$ arising from spin noise do {\em not} shift the X$^{1-}$ resonance backwards and forwards. Instead, a typical $B_{N}$ fluctuation induces a sub-linewidth Zeeman splitting of the X$^{1-}$ resonance, as shown in Fig.\ 3(d). An oscillatory $B_{N}$ results in a ``breathing motion" of the RF spectrum. Sensitivity to spin noise in the RF is therefore strong for $\langle \delta\rangle=0$, weak for $\langle \delta\rangle=\Gamma/2$. The crucial point is that, for X$^{1-}$ at $B_{\rm ext}=0$, the dependence of the RF noise on $\delta$ is {\em opposite} for charge noise and spin noise.

Applying this concept to the quantum dot response in Fig.\ 2(a) leads to the unambiguous conclusion that the noise at low frequencies arises from charge noise and that the noise at high frequencies arises from spin noise. The noise spectrum at $\langle \delta\rangle=0$ measured on an empty quantum dot, driving the neutral exciton X$^{0}$ transition, also shows two noise features, again charge noise and spin noise, Fig.\ 4(a). The X$^{0}$ and X$^{1-}$ have similar levels of charge noise. This is expected as the X$^{0}$ and X$^{1-}$ dc Stark shifts are similar and each exciton probes exactly the same environment. The X$^{0}$ spin noise is less however. Qualitatively, this reflects the fact that the X$^{0}$ splits into two states even at $B_{\rm ext}=0$ (the so-called fine structure, a consequence of an anisotropy in the electron-hole exchange) such that the dispersion for small $B_{N}$ is quadratic and not linear, reducing massively the sensitivity of X$^{0}$ to spin noise (see Methods).

The noise behaviour X$^{0}$ versus X$^{1-}$ supports the charge/spin assignment of the noise processes. Further confirmation is provided in Fig.\ 2(b) which shows $N_{\rm QD}(f)$ curves measured on the same quantum dot over the course of the experiment (several months) under nominally identical conditions. There are changes in the amplitude of the low frequency noise (up to a factor of 10) but the high frequency noise remains exactly the same. It is known that the charge state of the sample can change depending on the sample's history: these changes result in small shifts of the optical resonance. As shown here, these charge rearrangements result in changes in charge noise at low frequency. The spin noise arises mostly from the host nuclear spins of the quantum dot which remain the same and retain their properties: this results in the unchanging spin noise at high frequency.\\
\begin{figure}[t]  
\center
\includegraphics[width=\linewidth]{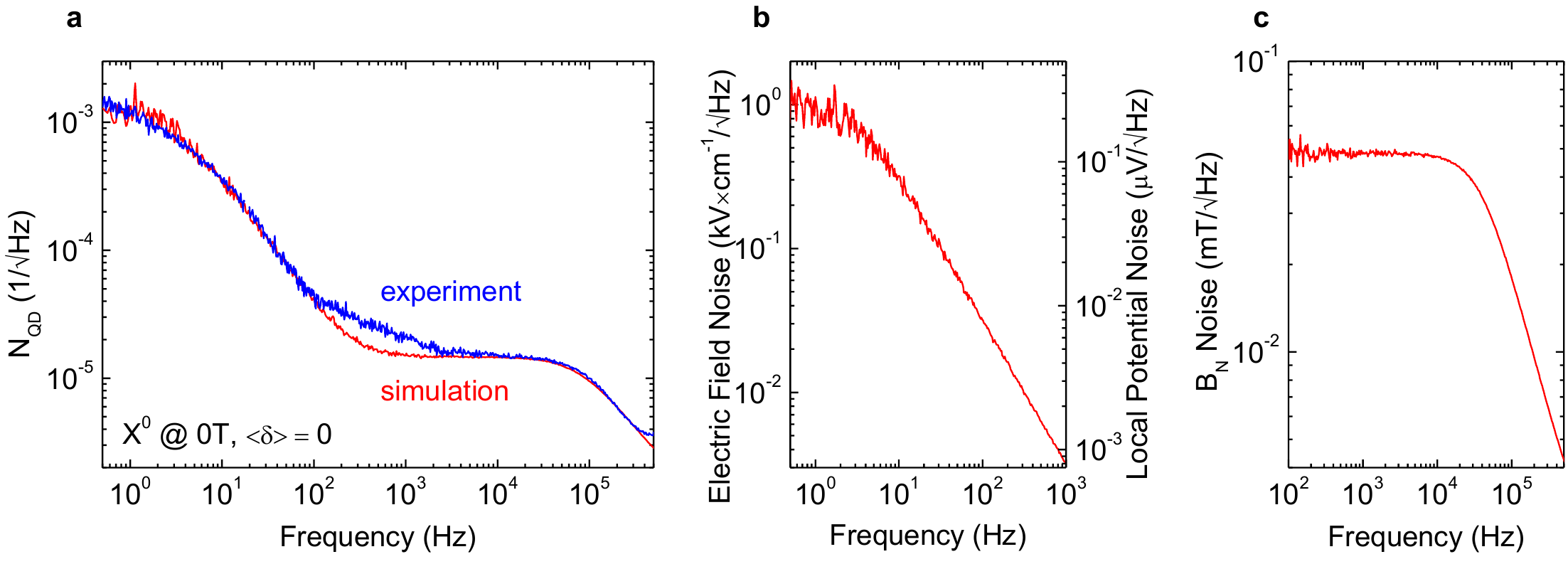}
\caption{{\bf Noise spectra of local electric and magnetic fields} (a) Experimental RF noise spectrum (blue) recorded on the neutral exciton X$^{0}$ (same quantum dot as in Fig.\ 1 and 2) with result of simulation (red). The simulation uses parameters $a=31.7$ $\mu$eV$\cdot$cm/kV, $N_C=0.2\times10^{10}$ cm$^{-2}$, $\tau_0 = 3$ s, $\tau_1=0.05 $ s, $p=1.6$ \% to model charge noise and $g = 0.5$, $\Delta=17.3$ $\mu$eV, $N = 375$, $\tau_0=\tau_1=8$ $\mu$s for spin noise. (b) Local electric field noise (left axis), local potential noise (right axis), (c) local magnetic field noise deduced from the simulations of the RF noise in (a).}
\end{figure}

\noindent{\bf Noise levels} Once the noise sources have been identified, the simple rules (see Methods) connecting RF intensity with the local electric field $F$ (charge noise) and with the local magnetic field $B_{N}$ (spin noise) allow quantitative statements on the noise to be made. The root-mean-square (rms) noise in the electric field is $F_{\rm rms}= 0.83$ Vcm$^{-1}$ with a characteristic frequency 20 Hz (correlation time 0.05 s). It is striking that, first, the charge noise is very small -- the rms noise in the local potential is just 2.1 $\mu$V. This is a consequence of both the ultra-pure material (samples of poorer quality exhibit more charge noise) and also the carefully controlled experimental conditions. The sensitivity of the quantum dot to the small levels of charge noise reflects the large Stark shifts and, on the one hand, the potential of quantum dots as ultra-sensitive electrometers\cite{Alen2003,Vamivakas2011,Houel2012}; and, on the other hand, the difficulty in generating transform-limited single photons from individual quantum dots. Secondly, it is striking that, even at 4 K, the charge noise is concentrated at such low frequencies.

The rms noise in the Overhauser field as measured on X$^{0}$ amounts to $B_{N,\rm rms} \simeq 20-40$ mT with a characteristic frequency 125 kHz (correlation time 8 $\mu$s). The prediction for the random fluctuations among $N$ independent nuclear spins is $B_{N,\rm rms}=A/\sqrt{N}$ where $A$ is the hyperfine coupling constant\cite{Merkulov2002,Khaetskii2002}. For $A \sim 90$ $\mu$eV, an average value for InGaAs\cite{Coish2009,Kloeffel2011}, and $N \sim 10^{5}$, $B_{N,\rm rms}\sim 10$ mT, reasonably close to the value measured here. The correlation time represents the time over which the nuclear spin distribution retains its memory. This noise-based measurement technique allows its direct determination. The time is comparable to estimates of nuclear spin decay through a dipole-dipole process\cite{Braun2005}. This direct observation of spin noise opens a new route to probing and controlling the nuclear spin evolution; experiments in a finite magnetic field are underway.\\

\noindent{\bf Quantum dot optical linewidth} A clear result is that both charge and spin noise fall rapidly with increasing frequency such that above 100 kHz, the RF noise amplitude (power) reduces by 2 to 3 (4 to 6) four orders of magnitude compared to the low frequency limit. The noise curves predict therefore that the exciton dephasing processes are slow relative to radiative decay which occurs at a GHz rate. To explore this, we measure the quantum dot linewidth as the measurement frequency $f_{\rm scan}$ is gradually increased (see Methods). Fig.\ 2(d) shows the RF linewidth $\Gamma$ as a function of scanning frequency. We find that $\Gamma$ decreases from 1.70 $\mu$eV to 0.93 $\mu$eV as $f_{\rm scan}$ increases from 1 Hz to 50 kHz. At higher $f_{\rm scan}$, $\Gamma$ remains constant. Furthermore, within our experimental error (0.1 $\mu$eV), this constant value at high $f_{\rm scan}$ corresponds to the transform limit, $\Gamma_{0}$. A transform-limited RF spectrum is shown in Fig.\ 2(c). In other words, the increase in $\Gamma$ over $\Gamma_{0}$ at low $f_{\rm scan}$ reflects the influence of processes which are slow not just relative to the recombination rate (GHz) but also relative to our maximum experimental ``speed" (100 kHz). These results are confirmed by measuring also X$^{1-}$ with the same procedure. In this case, the linewidth decreases to 0.75 $\mu$eV at high scan rates, a lower value than for X$^{0}$, reflecting the slightly larger radiative decay time\cite{Dalgarno2008a} for X$^{1-}$.\\

\noindent{\bf Charge noise and spin noise spectra} The local charge noise spectrum Fig.\ 2(a) has a roll-off behaviour (roughly constant at low frequencies, decreasing at higher frequencies). The charge noise power spectrum $[N_{\rm QD}(f)]^2$ can be well fitted by a Lorentzian function, characteristic of a two-level fluctuator\cite{Machlup1954}. The close-to-ubiquitous $1/f$ ``flicker" noise \cite{Weissman1988} is {\em not} observed.

A single two-level fluctuator would lead to telegraph noise in the RF which we do not observe in this experiment. Instead, we postulate that the noise arises from fluctuations in an {\em ensemble} of two-level fluctuators, each with approximately the same transition rates, $0 \rightarrow 1$, $1 \rightarrow 0$. The particular fluctuators are charge localization centres located close to the quantum dots: an additional elementary charge 150 nm above from the quantum dot shifts the resonance by a few $\mu$eV through the dc Stark effect\cite{Houel2012}. These fluctuating charges are holes in this case: surplus electrons relax rapidly into the Fermi sea whereas surplus holes, minority carriers, can be trapped in the active part of the device, the holes arising from the weak residual doping in the GaAs. Electrostatic noise arises on account of fluctuations in the exact configuration of occupied (state 0) and unoccupied (state 1) localization sites in the ensemble. We simulate the noise by taking (i) a fixed array of localization centres, (ii) a fixed average hole concentration, and (iii) a centre-independent capture/escape rate. The total electric field at the location of the quantum dot is calculated for a particular configuration from simple electrostatics\cite{Houel2012}, and the configuration changes over time are calculated by treating each localization centre separately in a Monte Carlo simulation (see Methods). 

The spin noise is modelled in a similar way, by treating each nuclear spin as a fictitious two-level system (see Methods). The simulations yield time traces $F(t)$ and $B_{N}(t)$. The RF signal $S(t)$ is then calculated according to the known dependence of RF on $F$ and $B_{N}$ (see Methods), and then a simulated noise amplitude $N_{\rm QD}(f)$ is calculated using exactly the same routine used to process the experimental data. The complete simulation accounts for simultaneous $F$ and $B_{N}$ fluctuations; it allows us to draw precise conclusions on the charge and spin noise without assuming for instance an over-simplified dependence of RF on $F$, $B_{N}$; and it enables us to perform a stringent test of the specific charge noise model.

The result of the simulation is shown in Fig.\ 4(a) where very close correspondence with the measured noise spectrum has been achieved. The amplitude of the low frequency noise, the charge noise, depends sensitively on the number, location and occupation probability of the localization centres; the characteristic roll-off frequency on the capture/escape rates. The amplitude of the high frequency noise, the spin noise, depends sensitively on $B_{N,\rm rms}$; its associated characteristic frequency depends on the nuclear spin flip rate. The success of the simulation allows us to present the noise spectra of $F$ and $B_{N}$ individually, Fig.\ 4 (b),(c): these curves are deduced from the measured RF noise with low random and systematic error.\\

\noindent{\bf Sample history} It is known that the optical resonance frequency of a particular quantum dot varies slightly from cool down to cool down. Fig.\ 2(b) shows in addition that the charge noise at low temperature is dependent on the sample's history. The amplitude of the low frequency noise varies by up to a factor of ten depending on the particular charge state of the sample. For this particular sample, the low noise state can be reached by temporary illumination with non-resonant laser light, followed by a wait of a few hours during which the noise at very low frequencies gradually reduces. This information is crucial in optimizing performance of the device as spin or optical qubit. The point we stress is that the noise spectrum is much more revealing about the dephasing processes than the optical frequency or optical linewidth alone.\\

\noindent{\bf Role of non-resonant excitation} The RF experiment involves driving the optical resonance with coherent laser light at photon energies far below the band gap of the host semiconductor. We have checked that the laser itself does not induce noise by carrying out experiments with two lasers, one tuned to the quantum dot resonance, the other detuned by a few nanometers. The presence of the second laser does not change the quantum dot noise spectrum. The situation changes profoundly if RF is detected in the presence of a second laser with photon energy above the band gap, non-resonant excitation. Even very small non-resonant intensities result in much increased noise. Initially, as the non-resonant power is increased, there is a rapid increase in noise at low frequencies, Fig.\ 5(a). Even measured slowly with 0.1 s integration time per point, there are massive changes in the RF, and, as a consequence, large changes in the exact lineshape from scan to scan, Fig.\ 5(b). On increasing the non-resonant power, this low frequency noise goes away -- the noise at the lowest frequencies returns almost to its original level -- but noise now appears at higher frequencies, Fig.\ 5(c), in particular between 10 Hz and 10 kHz. Measured slowly, the spectrum acquires a Lorentzian-shape without scan-to-scan fluctuations, Fig.\ 5(b), but with an increased linewidth as a consequence of the extra noise at frequencies above 10 Hz. At these non-resonant powers, the photoluminescence  induced by the non-resonant laser is {\em weaker} than the RF induced by the resonant laser. At higher non-resonant laser powers, the photoluminescence dominates over the RF and the noise increases further. These results demonstrate that while non-resonant illumination can change and possibly reduce fluctuations at low-frequency, it results in a {\em net increase} in noise. The standard optical technique, detection of photoluminescence with non-resonant excitation, has this serious flaw, expressed quantitatively in this experiment.
\begin{figure}[t]  
\center
\includegraphics[width=\linewidth]{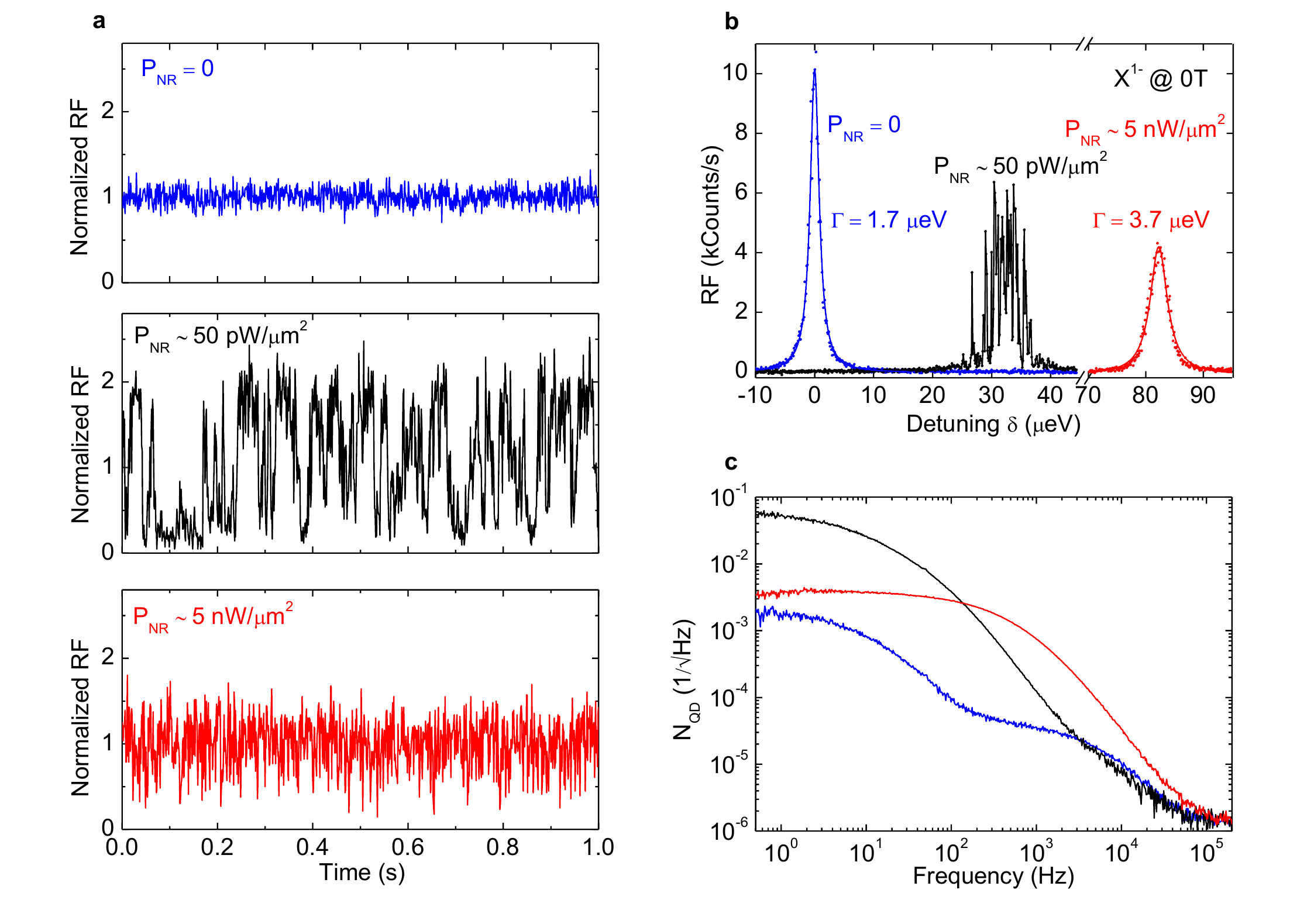}
\caption{{\bf Noise and above band gap excitation} (a) Normalized RF time traces from a single quantum dot (different quantum dot to Fig.s\ 1 and 2), (b) RF spectra (0.1 s integration per point), and (c) noise spectra plotted for X$^{1-}$ non-resonant power zero (blue), 1.3 nW (black) and 168 nW (red) focused to a spot area of $\sim 20$ $\mu$m$^{2}$. The non-resonant excitation induces initially considerable noise at low frequencies; larger non-resonant excitation sees the low frequency noise return close to the value observed without non-resonant excitation but considerable noise now appears above a few tens of Hz.}
\end{figure}

As an outlook, we comment that (i) the high frequency limit of our experiment is limited only by the photon flux. Which can be increased relatively simply using either a micro-cavity or photonic nanowire to enhance the photon extraction efficiency from the device. Our technique is potentially capable of mapping the noise from sub-Hz frequencies up to the MHz regime, possibly the GHz regime. (ii) The charge noise is measured here in a simple device without any post-growth micro- or nano-structuring: it represents a baseline for the local charge noise in an ultra-pure semiconductor. Furthermore, our quantitative description of the charge noise can be translated to other samples once the key parameters have been established. (iii) The sensitivity of the technique to spin noise opens a new route to probing the spin noise. Its dependence on external magnetic field, charge state of the quantum dot, laser excitation etc.\ can all be probed simply by recording time traces of the RF. (iv) The experiment demonstrates that the dephasing processes which limit the $T_{2}^{*}$ of the quantum dot exciton are all slow with respect to radiative recombination; that charge and spin noise reduce rapidly for increasing frequencies: the ``semiconductor vacuum" is achieved for the exciton above 50 kHz, and perhaps for a spin qubit above 1 MHz. These results all point to the possibilities of achieving close to dephasing-free qubit operations by working at very high frequencies or at lower frequencies by exploiting echo-like schemes.\\ 

\newpage \noindent {\bf Methods}\\
\small
\noindent {\bf Resonance fluorescence on a single quantum dot}\\
The sample consists of back contact (50 nm n$^{+}$-GaAs), tunneling barrier (25 nm i-GaAs ), InGaAs self-assembled quantum dots with centre wavelength 950 nm, capping layer (150 nm i-GaAs ), a blocking barrier (68 periods AlAs/GaAs 3 nm/1 nm), 10 nm GaAs, and a Schottky gate (3 nm/7nm Ti/Au). The background doping of as-grown GaAs is $p \sim 10^{13}$ cm$^{-3}$; two-dimensional electron gases grown under similar conditions have mobilities $> 10^{6}$ cm$^{2}$/Vs. The quantum dot optical resonance is driven with a resonant laser (1 MHz linewidth) focused on to the sample surface. Reflected or scattered laser light is rejected with a dark field technique using crossed linear polarizations for excitation and detection\cite{Yilmaz2010,Matthiesen2012,Houel2012}. The laser power is chosen to lie well below the point at which power broadening can be observed. Resonance fluorescence is detected with a silicon avalanche photodiode in photon counting mode. The arrival time of each photon is recorded over the entire measurement time $T$.\\ 

\noindent{\bf Determination of quantum dot RF noise spectrum}\\
Post measurement, a binning time $t_{\rm bin}$ is selected, typically 1 $\mu$s. The number of counts in each time bin is $S(t)$, the average number of counts per bin $\langle S(t)\rangle$. The fast Fourier transform of the normalized RF signal $S(t)/\langle S(t)\rangle$ is calculated to yield a noise spectrum, specifically $N_{\rm RF}(f)=\left| {\rm FFT}\left[S(t)/\langle S(t)\rangle\right]\right| t_{\rm bin}/\sqrt{T}$. $N_{\rm RF}(f)$ has the same spectrum independent of the choice of $t_{\rm bin}$ and $T$: smaller values of $t_{\rm bin}$ allow $N(f)$ to be determined to higher values of frequency $f$; larger values of $T$ allow $N(f)$ to be determined with higher resolution. To record a noise spectrum of the experiment alone, the quantum dot is detuned by $>100$ linewidths relative to the laser and one polarizer is rotated by a small angle to open slightly the detection channel for reflected laser light, choosing the rotation so that the detected laser light gives the same count rate as the quantum dot RF. A noise spectrum of the reflected laser light is recorded using the routine used to analyse the RF, yielding $N_{\rm exp}(f)$. $N_{\rm exp}(f)$ has a $1/f$-behaviour at low frequencies arising from intensity fluctuations in the laser: $N_{\rm exp}(f)$ is typically larger than $N_{\rm QD}(f)$ for $f<1$ Hz. For $f>100$ Hz, $N_{\rm exp}(f)$ has a completely $f$-independent spectrum, $N_{\rm exp} \sim 10^{-3}$ Hz$^{-1/2}$: this is the shot noise. The noise spectrum of the quantum dot is then determined using $N_{\rm QD}(f)=N_{\rm RF}(f)-N_{\rm exp}(f)$. Correction of $N_{\rm RF}(f)$ with $N_{\rm exp}(f)$ where $N_{\rm RF}(f)$ and $N_{\rm exp}(f)$ are not measured simultaneously is successful on account of the high stability of the setup. Furthermore, no spectral resonances in $N_{\rm QD}(f)$ have been discovered. We present here $N_{\rm QD}(f)$ after averaging at each $f$ over a frequency range $\Delta f$ yielding equidistant data points on a logarithmic scale. This entire procedure enables us to  discern $N_{\rm QD}(f)$ down to values of $10^{-5}$ Hz$^{-1/2}$ for $T=2$ hours.\\

\noindent{\bf Determination of quantum dot linewidth}\\
To determine the quantum dot optical linewidth $\Gamma$, we apply a triangle voltage signal to the gate with 100 mV amplitude, recording the RF signal as a function of time. Each time the quantum dot comes into resonance with the constant frequency laser, a peak in the RF is observed. The peak is fitted to a Lorentzian, and the width in mV is converted to a width in $\mu$eV using the Stark shift. The Stark shift is determined by recording the resonance position in $V_{g}$ for many laser frequencies, the laser frequency measured in each case with an ultra-precise wavemeter. The scan frequency is defined as the scanning rate divided by the transform-limited linewidth, $f_{\rm scan}=d\delta/dt/\Gamma_{0}$ with $\Gamma_{0}=\hbar/\tau_{r}$. The radiative lifetime, $\tau_{r}$, is measured either from an intensity correlation measurement, $g^{(2)}(t)$, or from a decay curve following pulsed excitation.\\

\noindent{\bf Noise simulations}
The experiment determines the spectrum of the noise in the RF and demonstrates that it is dominated by charge noise at low frequency, spin noise at high frequency. The noise sensor, the RF from a single quantum dot, has a trivial dependence on the fluctuating electric and magnetic fields only for small fluctuations in the detunings around particular values of detuning $\delta$, Fig.\ 3. The Monte Carlo simulations allow us to determine both the electric field and magnetic field noise accurately by describing the response of the sensor for all $\delta$, treating charge noise and spin noise on an equal footing. The basic approach is to calculate $F(t)$ and $B_{N}(t)$, in each case from a ensemble of independent 2-level fluctuators using a Monte Carlo approach; to calculate the RF signal $S(t)$ from $F(t)$ and $B_{N}(t)$; and to compute the noise $N(f)$ from $S(t)$ using exactly the same routine as for the experiments (but without the correction for extrinsic noise of course). For X$^{1-}$,
\begin{equation}
S=\frac{\frac{1}{2}\left(\frac{\Gamma_{0}}{2}\right)^{2}}{\left(aF+\delta_{1}\right)^{2}+\left(\frac{\Gamma_{0}}{2}\right)^{2}}+ 
\frac{\frac{1}{2}\left(\frac{\Gamma_{0}}{2}\right)^{2}}{\left(aF-\delta_{1}\right)^{2}+\left(\frac{\Gamma_{0}}{2}\right)^{2}},\; \; \; \delta_{1}=\frac{1}{2}g\mu_{B}B_{N},
\end{equation}
where $a$ is the dc Stark coefficient and $g$ the electron g-factor. For X$^{0}$, 
\begin{equation}
S=\frac{\left(\frac{\Gamma_{0}}{2}\right)^{2}}{\left(aF+\delta_{0}\right)^{2}+\left(\frac{\Gamma_{0}}{2}\right)^{2}},\; \; \;  \delta_{0}=\pm \frac{1}{2}\sqrt{\Delta^{2}+\delta_{1}^{2}},
\end{equation}
with $\Delta$ the fine structure splitting. 

The simulation for the charge noise proceeds by assuming that the noise arises from an ensemble of localization centres, each of which can be occupied by a single hole\cite{Houel2012}. The centres are located 150 nm away from the quantum dots at the capping layer/blocking barrier interface. The centres have a density of $N_{c}$ and, at any particular time, are occupied/unoccupied (states 1/0) with probability $p$, $1-p$ such that the hole density is $N_{h}=p N_{c}$. At $t=0$, each centre is occupied by a random number generator giving a configuration of localized charges $C(0)$. At a later time, $\delta t$, $C(\delta t)$ is calculated from $C(0)$ again with a random number generator using the probabilities of a $0 \rightarrow 1$, $1 \rightarrow 0$ transition from the theory of a two-level fluctuator\cite{Machlup1954}. The localization centres are treated independently. The localization centres directly above the quantum dot give rise to substantial energy shifts which we very rarely observe: we therefore neglect any localization centres in a circle of diameter 1 $\mu$m about the quantum dot axis. This is probably related to a strain field above the quantum dot. The procedure is repeated to give $C(0), C(\delta t), C(2 \delta t)$, etc. The electric field $F(t)$ is calculated\cite{Houel2012} for each $C$. The calculation of the time trace of the magnetic field $B_{N}$ proceeds in a similar way, albeit simplified: each nucleus is treated as a two-level fluctuator, with equal $0 \rightarrow 1$, $1 \rightarrow 0$ transition rates. \\

\noindent{\bf Acknowledgements}\\
We acknowledge financial support from NCCR QSIT. We thank Bill Coish, Christoph Kloeffel and Daniel Loss for helpful discussions; Sascha Martin and Michael Steinacher for technical support. A.L., D.R. and A.D.W. acknowledge gratefully support from DFG SPP1285 and BMBF QuaHLRep 01BQ1035.


\normalsize

\begin{thebibliography}{10}
\expandafter\ifx\csname url\endcsname\relax
  \def\url#1{\texttt{#1}}\fi
\expandafter\ifx\csname urlprefix\endcsname\relax\def\urlprefix{URL }\fi
\providecommand{\bibinfo}[2]{#2}
\providecommand{\eprint}[2][]{\url{#2}}

\bibitem{Loss1998}
\bibinfo{author}{Loss, D.} \& \bibinfo{author}{DiVincenzo, D.~P.}
\newblock \bibinfo{title}{Quantum computation with quantum dots}.
\newblock \emph{\bibinfo{journal}{Phys. Rev. A}} \textbf{\bibinfo{volume}{57}},
  \bibinfo{pages}{120--126} (\bibinfo{year}{1998}).

\bibitem{Petta2005}
\bibinfo{author}{Petta, J.~R.} \emph{et~al.}
\newblock \bibinfo{title}{{Coherent manipulation of coupled electron spins in
  semiconductor quantum dots}}.
\newblock \emph{\bibinfo{journal}{{Science}}} \textbf{\bibinfo{volume}{{309}}},
  \bibinfo{pages}{{2180--2184}} (\bibinfo{year}{{2005}}).

\bibitem{Shields2007}
\bibinfo{author}{Shields, A.~J.}
\newblock \bibinfo{title}{{Semiconductor quantum light sources}}.
\newblock \emph{\bibinfo{journal}{{Nature Photonics}}}
  \textbf{\bibinfo{volume}{{1}}}, \bibinfo{pages}{{215--223}}
  (\bibinfo{year}{{2007}}).

\bibitem{Fischer2009b}
\bibinfo{author}{Fischer, J.} \& \bibinfo{author}{Loss, D.}
\newblock \bibinfo{title}{{Dealing with Decoherence}}.
\newblock \emph{\bibinfo{journal}{{Science}}} \textbf{\bibinfo{volume}{{324}}},
  \bibinfo{pages}{{1277--1278}} (\bibinfo{year}{{2009}}).

\bibitem{Hogele2004}
\bibinfo{author}{H\"ogele, A.} \emph{et~al.}
\newblock \bibinfo{title}{{Voltage-controlled optics of a quantum dot}}.
\newblock \emph{\bibinfo{journal}{{Phys. Rev. Lett.}}}
  \textbf{\bibinfo{volume}{{93}}}, \bibinfo{pages}{{217401}}
  (\bibinfo{year}{{2004}}).

\bibitem{Atature2006}
\bibinfo{author}{Atat\"ure, M.} \emph{et~al.}
\newblock \bibinfo{title}{{Quantum-dot spin-state preparation with near-unity
  fidelity}}.
\newblock \emph{\bibinfo{journal}{{Science}}} \textbf{\bibinfo{volume}{{312}}},
  \bibinfo{pages}{{551--553}} (\bibinfo{year}{{2006}}).

\bibitem{Houel2012}
\bibinfo{author}{Houel, J.} \emph{et~al.}
\newblock \bibinfo{title}{Probing single-charge fluctuations at a GaAs/AlAs
  interface using laser spectroscopy on a nearby InGaAs quantum dot}.
\newblock \emph{\bibinfo{journal}{Phys. Rev. Lett.}}
  \textbf{\bibinfo{volume}{108}}, \bibinfo{pages}{107401}
  (\bibinfo{year}{2012}).

\bibitem{Klotz2010}
\bibinfo{author}{Klotz, F.} \emph{et~al.}
\newblock \bibinfo{title}{{Observation of an electrically tunable exciton g
  factor in InGaAs/GaAs quantum dots}}.
\newblock \emph{\bibinfo{journal}{{Appl. Phys. Lett.}}}
  \textbf{\bibinfo{volume}{{96}}}, \bibinfo{pages}{{053113}}
  (\bibinfo{year}{{2010}}).

\bibitem{Pingenot2011}
\bibinfo{author}{Pingenot, J.}, \bibinfo{author}{Pryor, C.~E.} \&
  \bibinfo{author}{Flatte, M.~E.}
\newblock \bibinfo{title}{{Electric-field manipulation of the Lande g tensor of
  a hole in an In(0.5)Ga(0.5)As/GaAs self-assembled quantum dot}}.
\newblock \emph{\bibinfo{journal}{{Phys. Rev. B}}}
  \textbf{\bibinfo{volume}{{84}}}, \bibinfo{pages}{{195403}}
  (\bibinfo{year}{{2011}}).

\bibitem{Merkulov2002}
\bibinfo{author}{Merkulov, I.~A.}, \bibinfo{author}{Efros, A.~L.} \&
  \bibinfo{author}{Rosen, M.}
\newblock \bibinfo{title}{{Electron spin relaxation by nuclei in semiconductor
  quantum dots}}.
\newblock \emph{\bibinfo{journal}{{Phys. Rev. B}}}
  \textbf{\bibinfo{volume}{{65}}}, \bibinfo{pages}{{205309}}
  (\bibinfo{year}{{2002}}).

\bibitem{Khaetskii2002}
\bibinfo{author}{Khaetskii, A.~V.}, \bibinfo{author}{Loss, D.} \&
  \bibinfo{author}{Glazman, L.}
\newblock \bibinfo{title}{Electron spin decoherence in quantum dots due to
  interaction with nuclei}.
\newblock \emph{\bibinfo{journal}{Phys. Rev. Lett.}}
  \textbf{\bibinfo{volume}{88}}, \bibinfo{pages}{186802}
  (\bibinfo{year}{2002}).

\bibitem{Greilich2006a}
\bibinfo{author}{Greilich, A.} \emph{et~al.}
\newblock \bibinfo{title}{{Mode locking of electron spin coherences in singly
  charged quantum dots}}.
\newblock \emph{\bibinfo{journal}{{Science}}} \textbf{\bibinfo{volume}{{313}}},
  \bibinfo{pages}{{341--345}} (\bibinfo{year}{{2006}}).

\bibitem{Xu2008}
\bibinfo{author}{Xu, X.} \emph{et~al.}
\newblock \bibinfo{title}{{Coherent population trapping of an electron spin in
  a single negatively charged quantum dot}}.
\newblock \emph{\bibinfo{journal}{{Nature Physics}}}
  \textbf{\bibinfo{volume}{{4}}}, \bibinfo{pages}{{692--695}}
  (\bibinfo{year}{{2008}}).

\bibitem{Press2010}
\bibinfo{author}{Press, D.} \emph{et~al.}
\newblock \bibinfo{title}{{Ultrafast optical spin echo in a single quantum
  dot}}.
\newblock \emph{\bibinfo{journal}{{Nature Photonics}}}
  \textbf{\bibinfo{volume}{{4}}}, \bibinfo{pages}{{367--370}}
  (\bibinfo{year}{{2010}}).

\bibitem{Balasubramanian2009}
\bibinfo{author}{Balasubramanian, G.} \emph{et~al.}
\newblock \bibinfo{title}{{Ultralong spin coherence time in isotopically
  engineered diamond}}.
\newblock \emph{\bibinfo{journal}{{Nature Materials}}}
  \textbf{\bibinfo{volume}{{8}}}, \bibinfo{pages}{{383--387}}
  (\bibinfo{year}{{2009}}).

\bibitem{Steger2012}
\bibinfo{author}{Steger, M.} \emph{et~al.}
\newblock \bibinfo{title}{{Quantum Information Storage for over 180 s Using
  Donor Spins in a Si-28 ``Semiconductor Vacuum{''}}}.
\newblock \emph{\bibinfo{journal}{{Science}}} \textbf{\bibinfo{volume}{{336}}},
  \bibinfo{pages}{{1280--1283}} (\bibinfo{year}{{2012}}).

\bibitem{Barthel2010}
\bibinfo{author}{Barthel, C.}, \bibinfo{author}{Medford, J.},
  \bibinfo{author}{Marcus, C.~M.}, \bibinfo{author}{Hanson, M.~P.} \&
  \bibinfo{author}{Gossard, A.~C.}
\newblock \bibinfo{title}{{Interlaced Dynamical Decoupling and Coherent
  Operation of a Singlet-Triplet Qubit}}.
\newblock \emph{\bibinfo{journal}{{Phys. Rev. Lett.}}}
  \textbf{\bibinfo{volume}{{105}}}, \bibinfo{pages}{{266808}}
  (\bibinfo{year}{{2010}}).

\bibitem{deLange2010}
\bibinfo{author}{de~Lange, G.}, \bibinfo{author}{Wang, Z.~H.},
  \bibinfo{author}{Riste, D.}, \bibinfo{author}{Dobrovitski, V.~V.} \&
  \bibinfo{author}{Hanson, R.}
\newblock \bibinfo{title}{{Universal Dynamical Decoupling of a Single
  Solid-State Spin from a Spin Bath}}.
\newblock \emph{\bibinfo{journal}{{Science}}} \textbf{\bibinfo{volume}{{330}}},
  \bibinfo{pages}{{60--63}} (\bibinfo{year}{{2010}}).

\bibitem{Bluhm2011}
\bibinfo{author}{Bluhm, H.} \emph{et~al.}
\newblock \bibinfo{title}{{Dephasing time of GaAs electron-spin qubits coupled
  to a nuclear bath exceeding 200 $\mu$s}}.
\newblock \emph{\bibinfo{journal}{{Nature Physics}}}
  \textbf{\bibinfo{volume}{{7}}}, \bibinfo{pages}{{109--113}}
  (\bibinfo{year}{{2011}}).

\bibitem{Coish2004}
\bibinfo{author}{Coish, W.~A.} \& \bibinfo{author}{Loss, D.}
\newblock \bibinfo{title}{{Hyperfine interaction in a quantum dot:
  Non-Markovian electron spin dynamics}}.
\newblock \emph{\bibinfo{journal}{{Phys. Rev. B}}}
  \textbf{\bibinfo{volume}{{70}}}, \bibinfo{pages}{{195340}}
  (\bibinfo{year}{{2004}}).

\bibitem{Urbaszek2013}
\bibinfo{author}{Urbaszek, B.} \emph{et~al.}
\newblock \bibinfo{title}{Nuclear spin physics in quantum dots: An optical
  investigation}.
\newblock \emph{\bibinfo{journal}{Rev. Mod. Phys.}}
  \textbf{\bibinfo{volume}{85}}, \bibinfo{pages}{79--133}
  (\bibinfo{year}{2013}).


\bibitem{Medford2012}
\bibinfo{author}{Medford, J.} \emph{et~al.}
\newblock \bibinfo{title}{Scaling of dynamical decoupling for spin qubits}.
\newblock \emph{\bibinfo{journal}{Phys. Rev. Lett.}}
  \textbf{\bibinfo{volume}{108}}, \bibinfo{pages}{086802}
  (\bibinfo{year}{2012}).

\bibitem{Reilly2008}
\bibinfo{author}{Reilly, D.~J.} \emph{et~al.}
\newblock \bibinfo{title}{Measurement of Temporal Correlations of the Overhauser Field in a Double Quantum Dot}.
\newblock \emph{\bibinfo{journal}{{Phys. Rev. Lett.}}} \textbf{\bibinfo{volume}{{101}}},
  \bibinfo{pages}{{236803}} (\bibinfo{year}{{2008}}).

\bibitem{Fink2013}
\bibinfo{author}{Fink, T.} \& \bibinfo{author}{Bluhm, H.}
\newblock \bibinfo{title}{Noise spectroscopy using correlations of single-shot
  qubit readout}.
\newblock \emph{\bibinfo{journal}{Phys. Rev. Lett.}}
  \textbf{\bibinfo{volume}{110}}, \bibinfo{pages}{010403}
  (\bibinfo{year}{2013}).

\bibitem{Drexler1994}
\bibinfo{author}{Drexler, H.}, \bibinfo{author}{Leonard, D.},
  \bibinfo{author}{Hansen, W.}, \bibinfo{author}{Kotthaus, J.~P.} \&
  \bibinfo{author}{Petroff, P.~M.}
\newblock \bibinfo{title}{{Spectroscopy of quantum levels in charge-tunable
  InGaAs quantum dots}}.
\newblock \emph{\bibinfo{journal}{{Phys. Rev. Lett.}}}
  \textbf{\bibinfo{volume}{{73}}}, \bibinfo{pages}{{2252--2255}}
  (\bibinfo{year}{{1994}}).

\bibitem{Warburton2000}
\bibinfo{author}{Warburton, R.~J.} \emph{et~al.}
\newblock \bibinfo{title}{{Optical emission from a charge-tunable quantum
  ring}}.
\newblock \emph{\bibinfo{journal}{{Nature}}} \textbf{\bibinfo{volume}{{405}}},
  \bibinfo{pages}{{926--929}} (\bibinfo{year}{{2000}}).

\bibitem{Alen2003}
\bibinfo{author}{Alen, B.}, \bibinfo{author}{Bickel, F.},
  \bibinfo{author}{Karrai, K.}, \bibinfo{author}{Warburton, R.~J.} \&
  \bibinfo{author}{Petroff, P.~M.}
\newblock \bibinfo{title}{{Stark-shift modulation absorption spectroscopy of
  single quantum dots}}.
\newblock \emph{\bibinfo{journal}{{Appl. Phys. Lett.}}}
  \textbf{\bibinfo{volume}{{83}}}, \bibinfo{pages}{{2235--2237}}
  (\bibinfo{year}{{2003}}).

\bibitem{Vamivakas2011}
\bibinfo{author}{Vamivakas, A.~N.} \emph{et~al.}
\newblock \bibinfo{title}{Nanoscale optical electrometer}.
\newblock \emph{\bibinfo{journal}{Phys. Rev. Lett.}}
  \textbf{\bibinfo{volume}{107}}, \bibinfo{pages}{166802}
  (\bibinfo{year}{2011}).

\bibitem{Coish2009}
\bibinfo{author}{Coish, W.~A.} \& \bibinfo{author}{Baugh, J.}
\newblock \bibinfo{title}{{Nuclear spins in nanostructures}}.
\newblock \emph{\bibinfo{journal}{{Phys. Stat. Sol. (b)}}}
  \textbf{\bibinfo{volume}{{246}}}, \bibinfo{pages}{{2203--2215}}
  (\bibinfo{year}{{2009}}).

\bibitem{Kloeffel2011}
\bibinfo{author}{Kloeffel, C.} \emph{et~al.}
\newblock \bibinfo{title}{{Controlling the Interaction of Electron and Nuclear
  Spins in a Tunnel-Coupled Quantum Dot}}.
\newblock \emph{\bibinfo{journal}{{Phys. Rev. Lett.}}}
  \textbf{\bibinfo{volume}{{106}}}, \bibinfo{pages}{{046802}}
  (\bibinfo{year}{{2011}}).

\bibitem{Braun2005}
\bibinfo{author}{Braun, P.-F.} \emph{et~al.}
\newblock \bibinfo{title}{{Direct Observation of the Electron Spin Relaxation Induced by Nuclei in Quantum Dots}}.
\newblock \emph{\bibinfo{journal}{{Phys. Rev. Lett.}}}
  \textbf{\bibinfo{volume}{{94}}}, \bibinfo{pages}{{116601}}
  (\bibinfo{year}{{2005}}).


\bibitem{Dalgarno2008a}
\bibinfo{author}{Dalgarno, P.~A.} \emph{et~al.}
\newblock \bibinfo{title}{{Coulomb interactions in single charged
  self-assembled quantum dots: Radiative lifetime and recombination energy}}.
\newblock \emph{\bibinfo{journal}{{Phys. Rev. B}}}
  \textbf{\bibinfo{volume}{{77}}}, \bibinfo{pages}{{245311}}
  (\bibinfo{year}{{2008}}).

\bibitem{Machlup1954}
\bibinfo{author}{Machlup, S.}
\newblock \bibinfo{title}{{Noise In Semiconductors - Spectrum of a 2-Parameter
  Random Signal}}.
\newblock \emph{\bibinfo{journal}{{J. Appl. Phys.}}}
  \textbf{\bibinfo{volume}{{25}}}, \bibinfo{pages}{{341--343}}
  (\bibinfo{year}{{1954}}).

\bibitem{Weissman1988}
\bibinfo{author}{Weissman, M.~B.}
\newblock \bibinfo{title}{{1/f noise and other slow, nonexponential kinetics in
  condensed matter}}.
\newblock \emph{\bibinfo{journal}{{Rev. Mod. Phys.}}}
  \textbf{\bibinfo{volume}{{60}}}, \bibinfo{pages}{{537--571}}
  (\bibinfo{year}{{1988}}).

\bibitem{Yilmaz2010}
\bibinfo{author}{Yilmaz, S.~T.}, \bibinfo{author}{Fallahi, P.} \&
  \bibinfo{author}{Imamoglu, A.}
\newblock \bibinfo{title}{{Quantum-Dot-Spin Single-Photon Interface}}.
\newblock \emph{\bibinfo{journal}{{Phys. Rev. Lett.}}}
  \textbf{\bibinfo{volume}{{105}}}, \bibinfo{pages}{{033601}}
  (\bibinfo{year}{{2010}}).

\bibitem{Matthiesen2012}
\bibinfo{author}{Matthiesen, C.}, \bibinfo{author}{Vamivakas, A.~N.} \&
  \bibinfo{author}{Atat\"ure, M.}
\newblock \bibinfo{title}{Subnatural linewidth single photons from a quantum
  dot}.
\newblock \emph{\bibinfo{journal}{Phys. Rev. Lett.}}
  \textbf{\bibinfo{volume}{108}}, \bibinfo{pages}{093602}
  (\bibinfo{year}{2012}).\\

\end{thebibliography}
\end{document}